\documentclass[twocolumn]{aastex6}
\usepackage{amssymb, amsmath,framed}

\usepackage{bm}
\expandafter\ifx\csname package@font\endcsname\relax\else
 \expandafter\expandafter
 \expandafter\usepackage
 \expandafter\expandafter
 \expandafter{\csname package@font\endcsname}
\fi
\def\bq{\begin{equation}}
\def\eq{\end{equation}}
\def\bqy{\begin{eqnarray}}
\def\eqy{\end{eqnarray}}

\def\p{\partial}

\def\rh{\rho}

\def\p{\partial}

\def\cale{\mathcal{E}}

\def\calm{\mathcal{M}}

\begin{document}
\title{Hall Current Effects in Mean-Field Dynamo Theory}

\author{Manasvi Lingam}
\email{mlingam@princeton.edu}
\affil{Department of Astrophysical Sciences and Princeton Plasma Physics Laboratory, Princeton University, Princeton, NJ 08544, USA}
\author{Amitava Bhattacharjee}
\affil{Department of Astrophysical Sciences and Princeton Plasma Physics Laboratory, Princeton University, Princeton, NJ 08544, USA \\
Max Planck-Princeton Center for Plasma Physics, Princeton University, Princeton, NJ 08544, USA}

\begin{abstract}
The role of the Hall term on large scale dynamo action is investigated by means of the First Order Smoothing Approximation. It is shown that the standard $\alpha$ coefficient is altered, and is zero when a specific double Beltrami state is attained, in contrast to the Alfv\'enic state for MHD dynamos. The $\beta$ coefficient is no longer positive definite, and thereby enables dynamo action even if $\alpha$-quenching were to operate. The similarities and differences with the (magnetic) shear-current effect are pointed out, and a mechanism that may be potentially responsible for $\beta < 0$ is advanced. The results are compared against previous studies, and their astrophysical relevance is also highlighted.
\end{abstract}

\maketitle

\section{Introduction} \label{SecIntro}
The importance of large scale magnetic fields in the Universe, ranging from the cosmic to the planetary, cannot be overstated \citep{KZ08,Sub16}. They play a chief role in processes such as star formation, galaxy evolution, etc. and as a means of probing and interpreting astrophysical environments. The principal process responsible for generating these large scale fields is the dynamo mechanism, which has been extensively studied for nearly a century \citep{KR80,YIIY04,BS05}, but there are several key issues that are yet unresolved \citep{BSS12}. 

To date, most analyses in plasma astrophysics, and, by extension, in dynamos, rely on ideal or resistive magnetohydrodynamics (MHD) as the underlying physical model on account of its simplicity. However, the MHD approximation is \emph{not} universally valid, and there are many domains where other plasma effects become dominant. One of the simplest of these `extended' models is Hall MHD, in which the ions and electrons generally move with different flow velocities. This effect is manifested in the Ohm's law via the Hall term, and the relative magnitude of such a term is typically encoded in the Hall parameter - the ratio of the ion skin depth to the characteristic background scale length.

In recent times, the Hall parameter has been shown to be `large' in diverse environments such as protoplanetary discs \citep{Ward07}, the ISM, the Earth's magnetotail \citep{Bhat04}, the solar corona and the solar wind \citep{Bhat04,GB07} to name a few. Furthermore, it is also likely to be highly relevant in space and laboratory settings \citep{Huba95,BMW01}, especially in experiments involving the Reversed Field Pinch \citep{Ji94,Ji99,Det04}. In each of these instances, understanding the dynamics of the magnetic fields would arguably necessitate the use of Hall MHD, as opposed to MHD, as the base physical model.  

Despite the ubiquity and importance of the Hall term, dynamo models that incorporate this term have been few and far between. The notable exceptions in large scale dynamo theory are the series of works undertaken by Mininni and collaborators \citep{MGM02,MGM03,MGM05,MAP07} as well as a few other publications \citep{Ji99,LM15}. There have also been some concomitant studies of small-scale Hall MHD dynamos, such as \citet{KR94,MHP03,GMD10}. The Hall MRI dynamo has been subjected to detailed investigations only in recent times \citep{SS02,KL13,LB16}.

In this paper, we shall use incompressible Hall MHD and carry out an investigation along the lines of \citet{GD94} and \citet{GD95}, which will henceforth be referred to as GD94 and GD95 respectively. At this stage, a crucial comment pertaining to our approach is necessary. Our model does not include stochasticity and chaos \citep{VB97,BT07}, the shear-current effect \citep{RK03,RK04,SB16}, helicity flux contributions \citep{VC01,SB04,SSS07,EB14}, dynamical equations for the large- and small-scale helicities \citep{BF02,BSS12,Bla15}, and hyperresistivity \citep{BY95,BS05} to name a few. Although each and every one of these effects is undoubtedly important, the central goal of the paper is to understand how the Hall term acts in near-isolation.

We shall show that the Hall term introduces several non-trivial effects, such as the existence of a non-positive definite diffusion coefficient, the drive towards a non-Alfv\'enic final state, and the possibility for dynamo action despite $\alpha$-quenching \citep{VC92,GD94,CH96} amongst others. Collectively, we argue that these non-trivial effects make a compelling case for carrying out in-depth Hall MHD dynamo analyses in the future. 

\section{Mathematical preliminaries and the kinematic picture} \label{SecMPRP}
We briefly outline the mathematical equations of our model, and present a kinematic derivation of the dynamo coefficients in this Section.

\subsection{The governing equations of the model} \label{SSecGovEq}
Let us commence our analysis by presenting the equations for incompressible Hall MHD wherein the density obeys the relation $\rho = \mathrm{const}$. The dynamical equation for the velocity is
\begin{equation} \label{MomEqn}
\left[\frac{\p}{\p t} + {\bf V} \cdot \nabla\right] {\bf V}  = \left(\nabla \times {\bf B}\right) \times {\bf B} - \nabla \left(\frac{p}{\rh} \right),
\end{equation}
and the Ohm's law can be expressed as
\begin{equation} \label{HOhm}
\frac{\p {\bf B}}{\p t}  = \nabla \times \left({\bf V}_E \times {\bf B}\right) - \eta \nabla \times {\bf J},
\end{equation}
where ${\bf J} = \nabla \times {\bf B}$, ${\bf V}_E = {\bf V} - d_i {\bf J}$, $\eta$ is the resistivity, and $d_i$ is the ion skin depth. Also note that these equations have been expressed in Alfv\'enic units where ${\bf B}$ has the units of (Alfv\'en) velocity. The above set of equations must be supplemented with $\nabla \cdot {\bf V} = \nabla \cdot {\bf B} = 0$.

Before proceeding further, an interesting mathematical property is worth emphasizing. If we let ${\bf B} \rightarrow - {\bf B}$ in (\ref{HOhm}), we find that the equation is \emph{not} invariant under this transformation. In other words, it is the Hall term (that is linear in $d_i$) that breaks this symmetry. Hence, we may expect all terms linear in $d_i$ that appear in the electromotive force to violate this symmetry - we shall establish the validity of this conjecture in Sec. \ref{SSecNKP}.\footnote{The value of such a symmetry analysis has already been well established in gauge theories \citep{Wein76,JDJ99}.}

\subsection{A kinematic study of the Hall dynamo}\label{SSecKinHD}
To maintain consistency with GD94 and GD95, we adopt the same properties of the velocity field, i.e. the large scale component is taken to be negligible. This is, of course, quite a strong simplification since the inclusion of large scale velocity with shear leads to the ``shear-current'' effect \citep{RK03,RK04,SB16}. We split the magnetic field into the mean field ${\bf B}_0$, and the small-scale component ${\bf b}$. The mean field evolves via
\begin{equation} \label{KinMFB}
\frac{\p {\bf B}_0}{\p t}  = \nabla \times \boldsymbol{\cale} - d_i \nabla \times \left({\bf J}_0 \times {\bf B}_0\right) - \eta \nabla \times {\bf J}_0,
\end{equation}
where the electromotive force $\boldsymbol{\cale}$ is determined through 
\begin{equation} \label{EMF}
\boldsymbol{\cale} = \langle{\left({\bf v} - d_i \nabla \times {\bf b}\right) \times {\bf b}\rangle},
\end{equation}
and $\langle{\dots\rangle}$ denotes the ensemble average (over the turbulent field ${\bf v}$). Upon applying the well-known first order smoothing approximation (FOSA) from kinematic dynamo theory \citep{KR80,BS05}, we find that the turbulent component of the magnetic field obeys
\begin{equation} \label{TurbKin}
\frac{\p {\bf b}}{\p t}  = \nabla \times \left[\left({\bf v} - d_i {\bf j}\right) \times {\bf B}_0\right] - d_i \nabla \times \left({\bf J}_0 \times {\bf b}\right).
\end{equation}
In obtaining the above relation, note that the resistive term has been neglected, as it does not play a major role in our subsequent analysis; for more details, refer to Section 6.3.1 of \citet{BS05}. Now, we can substitute (\ref{TurbKin}) into (\ref{EMF}) and implement the Reynolds relations, and the assumptions of statistical homogeneity and isotropy \citep{KR80,BS05}. The latter amounts to replacing correlation tensors (for turbulent quantities) of rank 2 and 3 with $\delta_{ij}$ and $\epsilon_{ijk}$ respectively. Upon doing so, we find that (\ref{EMF}) simplifies to
\begin{equation} \label{EMFAB}
\boldsymbol{\cale} =  \alpha {\bf B}_0 - \beta \nabla \times {\bf B}_0 + \dots,
\end{equation}
where the `$\dots$' indicate the existence of higher-order derivative terms in ${\bf B}_0$. 

Let us now turn our attention to (\ref{EMF}) in order to understand how it must be computed. We observe that it may be written as 
\begin{eqnarray} \label{EMFExp}
\boldsymbol{\cale} &=&  \tau_c \Big[\langle{\partial_t {\bf v} \times {\bf b}\rangle} + \langle{{\bf v} \times \partial_t {\bf b} \rangle} - d_i \langle{\left(\nabla \times {\bf b}\right) \times \partial_t {\bf b}\rangle} \nonumber \\
&&\quad - d_i \langle{\left(\nabla \times \partial_t {\bf b}\right) \times {\bf b}\rangle} \Big],
\end{eqnarray}
upon invoking the correlation time approximation \citep{BS05}; in the above expression, $\tau_c$ denotes the correlation time. It is easy to verify that the second and third terms in the above expression can be clubbed together as $\langle{{\bf v}_E \times \partial_t {\bf b}\rangle}$. We have dubbed the last three terms of (\ref{EMFExp}) as `kinematic' since they depend solely on the evolution of $\partial_t {\bf b}$, and are thus analogous to the standard kinematic treatments of the MHD dynamo \citep{BS05}. It is, however, important to recognize that there \emph{are} non-linear effects arising from the Hall drift, but these terms are not reliant upon the momentum equation (\ref{MomEqn}). On the other hand, the first term in (\ref{EMFExp}) is markedly different as it explicitly depends upon the temporal evolution of ${\bf v}$. For this reason, the first term is tackled in greater detail in the next section.

We shall concern ourselves presently with the kinematic picture by evaluating the last three terms in (\ref{EMFExp}). We find that the coefficients occurring in (\ref{EMFAB}) are thus given by
\begin{equation} \label{alphakin}
\alpha_0 = - \frac{\tau_c}{3} \Big\langle{{\bf v}_E \cdot \left(\nabla \times {\bf v}_E\right) + d_i {\bf b} \cdot \left(\nabla \times \nabla \times {\bf v}_E\right)\Big\rangle},   
\end{equation}
\begin{eqnarray} \label{betakin}
\beta_0 &=& \frac{\tau_c}{3} \Big\langle{\bf v}_E^2 + d_i \left({\bf b}\cdot \nabla \times {\bf v}_E + {\bf v}_E \cdot \nabla \times {\bf b} \right) \nonumber \\
&& \quad \quad +\, d_i^2\, {\bf b}\cdot \nabla \times \nabla \times {\bf b} \Big\rangle,  
\end{eqnarray}
where the subscript `$0$' associated with the coefficients indicates that they represent the kinematic values. In (\ref{alphakin}) and (\ref{betakin}), ${\bf v}_E = {\bf v} - d_i \nabla \times {\bf b}$ represents the (turbulent) electron velocity. The first and second terms in the $\alpha$-coefficient, whose explicit expression can be found in (\ref{alphakin}), arise from $\langle{{\bf v}_E \times \partial_t {\bf b}\rangle}$ and $\langle{\left(\nabla \times \partial_t {\bf b}\right) \times {\bf b}\rangle}$ respectively by means of (\ref{EMF}). Similarly, the first and second terms in (\ref{betakin}) originate via $\langle{{\bf v}_E \times \partial_t {\bf b}\rangle}$, and the last two terms in (\ref{betakin}) emerge from $\langle{\left(\nabla \times \partial_t {\bf b}\right) \times {\bf b}\rangle}$.

A few comments are in order. Firstly, we note that the \emph{dimensional} limit $d_i \rightarrow 0$ transforms (\ref{alphakin}) and (\ref{betakin}) to the standard $\alpha$ and $\beta$ contributions from MHD kinematic dynamo theory \citep{KR80}. Secondly, an inspection of (\ref{betakin}) reveals that $\beta_0$ is no longer positive definite - a feature that we explore in greater detail in Sec. \ref{SecImp}. Lastly, it is also possible to compute higher order contributions that appear in (\ref{EMFAB}). The so-called $\gamma$-effect \citep{MAP07} is existent when $\boldsymbol{\cale}$ contains terms that are proportional to $\nabla \times \left( \nabla \times {\bf B}_0\right)$. Upon implementing the same procedure, we find that
\begin{equation}
\gamma_0 = -\frac{\tau_c}{3} d_i\Big\langle{{\bf v}_E \cdot {\bf b}\Big\rangle},
\end{equation}
which clearly vanishes in the MHD limit, i.e. upon taking $d_i \rightarrow 0$.

\section{Beyond the kinematic paradigm} \label{SecNKD}
We tackle the nonlinear picture in this Section, and briefly outline alternative approaches to our methodology in conjunction with the accompanying results.

\subsection{The nonlinear regime of the Hall dynamo} \label{SSecNKP}
To tackle the `nonlinear' picture, we need to investigate the first term of (\ref{EMFExp}), as pointed out in Sec. \ref{SSecKinHD}. This term allows us to account for the back-reactions arising from the Lorentz force in the momentum equation.

At this stage, we point out a remarkable fact. Since the first term is given by $\langle{\partial_t {\bf v} \times {\bf b}\rangle}$, it can involve Hall corrections only through the quantity $\partial_t {\bf v}$ since there are no other factors of $d_i$ prevalent. However, it is well known that the evolution equation for the velocity (\ref{MomEqn}) is exactly \emph{identical} to its ideal MHD counterpart \citep{Huba95}. As a result, upon utilizing FOSA and the other assumptions of our model, it becomes Eq. (5) of GD94; see also Eq. (25) of GD95. The expression is given by
\begin{equation} \label{vEvolTurb}
    \partial_t {\bf v} = \left({\bf B}_0 \cdot \nabla\right){\bf b} + \left({\bf b} \cdot \nabla\right){\bf B}_0 - \nabla \left(\frac{p}{\rho}\right),
\end{equation}
and it turns out that the $\alpha$ correction arises only from the first term on the RHS in the above equation. This is easy to verify after some careful inspection, or by following through the analysis in GD94 and GD95. Upon plugging the first term on the RHS of (\ref{vEvolTurb}) into the first term on the RHS of (\ref{EMFExp}), we obtain the nonlinear correction. It turns out to be proportional to $\langle{{\bf b}\cdot \nabla \times {\bf b}\rangle}$, and thus the final expression for $\alpha$ is given by
\begin{eqnarray} \label{alphaNkin}
\alpha &=& - \frac{\tau_c}{3} \Big\langle{{\bf v}_E \cdot \left(\nabla \times {\bf v}_E\right) + d_i {\bf b} \cdot \left(\nabla \times \nabla \times {\bf v}_E\right)\Big\rangle} \nonumber \\
&& +  \frac{\tau_c}{3} \Big\langle{{\bf b}\cdot \left(\nabla \times {\bf b}\right)\Big\rangle},
\end{eqnarray}
and the term on the second line of the RHS is the nonlinear contribution arising from the back-reaction of the Lorentz force. As before, the limit $d_i \rightarrow 0$ leads to the classical nonlinear $\alpha$ coefficient first derived in \citet{PFL76}. It is instructive to compute how the nonlinear value of $\alpha$ differs from the kinematic value. In order to do so, we mirror the approach employed in GD94 and GD95. We write down the expression for the evolution of the vector potential (in a suitable gauge):
\begin{equation} \label{aevolEqn}
\partial_t {\bf a} = {\bf v}_E \times {\bf B}_0 + {\bf v}_E \times {\bf b} - \boldsymbol{\cale} - d_i {\bf J}_0 \times {\bf b} - \eta \nabla \times {\bf b},
\end{equation}
where ${\bf J}_0 = \nabla \times {\bf B}_0$ and $\boldsymbol{\cale}$ was defined in (\ref{EMF}). We take the dot product of the above equation with ${\bf b}$ and then perform the ensemble averaging. To simplify the resultant relation, we invoke the helicity balance (based on small-scale stationarity) argument of GD94 and GD95, and use the relations $\langle{{\bf b}\rangle} = 0$ and ${\bf b}\cdot \left({\bf b} \times {\bf X}\right) = 0$. We note that the above steps are justified since magnetic helicity is an invariant of Hall MHD, just as in ideal MHD \citep{Turn86,LMM15,LMM16}. Before proceeding further, we emphasize that the results of GD94 and GD95 are not truly universal, as they have been generalized by several authors - see for e.g. the work of \citet{BY95} on hyperresistivity and \citet{BF99,Black03,BS05}. 

After applying the aforementioned relations to (\ref{aevolEqn}) and carrying out the requisite algebra, we end up with
\begin{equation}
\langle{{\bf b} \cdot \left({\bf v}_E \times {\bf B}_0\right)\rangle} = \eta \langle{{\bf b}\cdot \nabla \times {\bf b}\rangle},
\end{equation}
and the LHS is further simplified by noting that $\langle{{\bf b} \cdot \left({\bf v}_E \times {\bf B}_0\right)\rangle} = - {\bf B}_0 \cdot \langle{{\bf v}_E \times {\bf b}\rangle} = - {\bf B}_0 \cdot \boldsymbol{\cale}$. The last equality followed upon using the definition of ${\bf v}_E$ in conjunction with (\ref{EMF}). Thus, we recover
\begin{equation} \label{CurHel}
    \langle{{\bf b}\cdot \nabla \times {\bf b}\rangle} = -\frac{\alpha}{\eta} B_0^2 + \frac{\beta}{\eta} {\bf B}_0 \cdot \nabla \times {\bf B}_0.
\end{equation}
Upon substituting (\ref{CurHel}) into (\ref{alphaNkin}), we find that
\begin{equation} \label{alphaquench}
\alpha = \left(\alpha_0 + \frac{\beta}{\eta} {\bf B}_0 \cdot \nabla \times {\bf B}_0\right) \left(1 + \frac{\tau_c B_0^2}{3 \eta}\right)^{-1},
\end{equation}
and it is possible to normalize the above equation in terms of the ``Zel'dovich'' units to obtain a result exactly analogous to GD94 and GD95. The only difference (thus far) is that the value of $\alpha_0$ is given by (\ref{alphakin}), which is different from its MHD counterpart. However, note that our analysis is not fully complete since the quantity $\beta$ in (\ref{alphaquench}) involves potential nonlinear contributions, which are yet to be evaluated.

Fortunately, we can take advantage of the aforementioned fact that the velocity evolution equation is the same for Hall and ideal MHD. We shall not present the full details of our calculation, and instead sketch the essential steps. The Fourier representation of the velocity and magnetic field evolution equations is considered, analogous to Appendix B in GD95. The second term in Eq. (B2) is the crucial quantity, as it represents the nonlinear contributions. As it involves $\partial_t {\bf v}$, the vertex $\Gamma_{\alpha \beta \gamma}$ in Eq. (B3) is identical, because of the above equivalence between Hall and ideal MHD. Consequently, the rest of the algebra, and the ensuing result, remains the same. Hence, we are led to the conclusion that the nonlinear contributions, in three dimensions, to the $\beta$ effect are nil. As a result, we can replace $\beta$ in (\ref{alphaquench}) with $\beta_0$, where the latter is given by (\ref{betakin}). 

Let us turn our attention to (\ref{alphaNkin}) and (\ref{betakin}) and perform the transformation ${\bf b} \rightarrow - {\bf b}$. We find that all terms linear in $d_i$ are rendered non-invariant, i.e. they change parity, thereby confirming the prediction stated in Sec. \ref{SSecGovEq}, and the value of this analysis.

\subsection{Alternative approaches to handling the Hall term} \label{SSecComp}
We commence by observing that (\ref{betakin}) and (\ref{alphaNkin}) represent our final dynamo coefficients. They have been obtained through a procedure analogous to the one outlined in GD94 and GD95. We also carried out the analysis for the Hall MHD large-scale dynamo using an alternative approach - a slightly modified version of the Minimal Tau Approximation (MTA) \citep{BF02,RR07}. We find that the same results, namely (\ref{betakin}) and (\ref{alphaNkin}), are recovered.

As noted in Sec. \ref{SecIntro}, Hall MHD large scale dynamos have not been studied widely in the literature. On the theoretical front, \citet{MGM02,MGM03} relied upon a variant of the Reduced Smoothing Approximation (RSA) \citep{BF99} to compute the $\alpha$-coefficient. We find that our result is in exact agreement with their work. The $\beta$-coefficient was computed later by \citet{MAP07} using the same approach. We have verified that the two expressions for $\beta$, namely (\ref{betakin}) and Eq. (3.10) of \citet{MAP07}, are equivalent upon integration by parts, up to an overall divergence term that is proportional to $\nabla \cdot \langle{{\bf b} \times \left(\nabla \times {\bf b}\right)}\rangle$. It can vanish in certain scenarios, for e.g. when (i) periodic boundary conditions exist, or (ii) the current exhibits a sufficiently steep decline and becomes zero at infinity (in the case of spatial averaging).

\section{On the implications of Hall MHD for large scale dynamos} \label{SecImp}
In this Section, we shall address the implications of the Hall term on large scale dynamos, and indicate the environments wherein such effects are likely to be important.

\subsection{The effect of the Hall term on the alpha coefficient}
We shall begin our study by considering the nonlinear expression for $\alpha$, which is given by (\ref{alphaNkin}). 

It has been argued that nonlinear Alfv\'en waves serve as the `building' blocks of plasma turbulence \citep{Bisk03,HN13}. In MHD, these waves obey the relation ${\bf b} = \pm {\bf v}$. In addition to the important property of equipartition, the Alfv\'enic solution has a crucial implication for large-scale dynamo theory - the $\alpha$-effect attains a null value \citep{PFL76,GD94}. The importance of Alfv\'enic states is further underscored by the fact that they are the asymptotic outcomes of MHD turbulence \citep{DMV80,GFPL82,Bol06}.

On the other hand, when we consider our dynamo model, we find that Alfv\'enic fluctuations do \emph{not} lead to $\alpha = 0$ due to the modifications induced by the Hall term. This condition is realized only when a very special ansatz is chosen -- the famous double Beltrami states of Hall MHD \citep{MY98}. It is obtained by extremizing the energy $E = \frac{1}{2} \int_D \left(v^2 + b^2\right)\,d^3x$ subject to the constraint that the two helicities
\begin{enumerate}
\item $H_1 =  \int_D {\bf a}\cdot {\bf b}\,d^3x$, which denotes the usual magnetic helicity
\item $H_2 = \int_D \left({\bf a} + d_i {\bf v}\right) \cdot \left({\bf b} + d_i \nabla \times {\bf v}\right)\,d^3x$, which is often referred to as the canonical, or generalized, helicity \citep{Turn86,LMM15,LMM16}
\end{enumerate}
are held constant. In mathematical terms, it is expressed via the variational principle $\delta F = 0$, where $F := E - 1/2\left(\lambda_1 H_1 + \lambda_2 H_2\right)$ is the relevant functional \citep{Turn86,MY98}. Upon simplification, the double Beltrami states of Hall MHD are given by
\begin{eqnarray} \label{DBelSolns}
{\bf v} &=& d_i \lambda_2 \left({\bf b} + d_i \nabla \times {\bf v}\right), \nonumber \\
\nabla \times {\bf b} &=& \lambda_1 {\bf b} + \lambda_2 \left({\bf b} + d_i \nabla \times {\bf v}\right).
\end{eqnarray}
For the multi-Beltrami states, it was also shown that $F = 0$ upon subsequent simplification \citep{ML15}. Thus, the resultant condition for $\alpha = 0$ is given by $\lambda_1 - \lambda_2 - 2 \lambda_2 \lambda_1^2 d_i^2 = 0$. As the condition $F = 0$ is also applicable for the double Beltrami states, we have two equations for $\lambda_1$ and $\lambda_2$. Hence, one can fully determine the precise values of $\lambda_1$ and $\lambda_2$, in terms of $E$, $H_1$, $H_2$ and $d_i$ when the $\alpha$-coefficient vanishes. For this specific double Beltrami state, i.e. when $\alpha$ is zero, we note that the condition $\beta > 0$ holds true.

In general, note that the double Beltrami solution has two free parameters. Owing to this freedom, it is possible for $\alpha$ to acquire positive, negative, or null values. This fact was also pointed out in \citet{MGM02}, but it is important to recognize that the fluctuations do not necessarily need to exhibit a double Beltrami behaviour. However, it is possible to verify that the system must attain a \emph{specific} double Beltrami state for the $\alpha$-coefficient to vanish. 

We have discussed the connection between Alfv\'en wave turbulence and the vanishing of the $\alpha$-coefficient in MHD. Next, we have considered Hall MHD and shown that a particular double Beltrami state is responsible for the vanishing of $\alpha$. Hence, it is natural to conjecture that the nonlinear Alfv\'en wave solutions of Hall MHD must be related (or can be constructed) from the double Beltrami states. This would ensure that Alfv\'en wave turbulence and large scale dynamo theory are equally connected in Hall MHD (as in MHD). We note that this intuitive picture is indeed correct, as the validity of the above conjecture was established in \citet{MK05}; see also Appendix B of \citet{LB16} for a discussion of the same. In mathematical terms, the Hall MHD nonlinear Alfv\'en waves obey the pair of relations
\begin{eqnarray} \label{NAHSol}
    {\bf v} - \nabla \times {\bf b} &=& \alpha {\bf b}, \\ \nonumber
 {\bf b} + \nabla \times {\bf v} &=& \frac{1}{\alpha} {\bf v},
\end{eqnarray}
which, upon suitable manipulation, can be reduced to a particular case of (\ref{DBelSolns}). The details of the algebra behind (\ref{NAHSol}) can be found in Sec. II of \citet{MK05}. Lastly, we wish to emphasize that the double Beltrami state, which leads to a null value of the $\alpha$-coefficient and shares connections with Hall MHD wave turbulence, is also accompanied by a lack of equipartition. The latter trait is quantified via the expression
\begin{equation} \label{BVNAHRel}
{\bf b}_k = \alpha_\pm {\bf v}_k,
\end{equation}
where the subscript `$k$' denotes the Fourier component of the corresponding fields. Here, $\alpha_\pm$ is given by
\begin{equation}
\alpha_\pm = - \frac{k}{2} \pm \sqrt{\frac{k^2}{4} + 1},
\end{equation}
and the wave vector $k$ has been normalized in units of $d_i$. An inspection of (\ref{BVNAHRel}) reveals clear differences upon comparison with the MHD Alfv\'enic solution ${\bf b} = \pm {\bf v}$, and it is also readily apparent that there is a clear non-equipartition; the latter arises because of the $k$-dependence that leads to $b_k^2 \neq v_k^2$. 

\subsection{The effect of the Hall term on the beta coefficient}
One of the important consequences of our work is that $\beta$ is \emph{not} positive definite, as opposed to its MHD counterpart. This feature has important consequences since it enables a non-local energy transfer from the small to large scales, as shown in the simulations by \citet{MAP07}. The analytical work presented in \citet{MAP07} also establishes that $\beta < 0$ is feasible, and agrees with our findings.

The shear-current effect that was analyzed in detail by \citet{RK03,RK04} has sparked a fair degree of controversy and extensive further research \citep{RK06,Betal08,Yet08,HP09,SS10}, but one of the appealing features of shear is that the $\beta$ effect can acquire negative off-diagonal contributions. The standard paradigm for the shear-current effect was recently revised by \citet{SB15,SB16}, who also coined the phrase magnetic shear-current effect. Through a combination of numerical and analytic work, these authors confirmed the existence of a negative anisotropic diffusivity.

However, our work is very different as it functions in the \emph{absence} of large-scale velocity shear, and we operate with a much simpler ansatz endowed with isotropy. It is important to recognize that the $\alpha$-quenching occurs via (\ref{alphaquench}), but $\beta$ is not subject to such a restriction. Thus, the fact that $\beta < 0$ is permitted within our Hall MHD dynamo enables the mechanism to be functional, even when $\alpha$ may be quenched. For this reason, it is instructive to see our work as fundamentally \emph{different}, but also \emph{complementary} to the above papers. In the latter scenarios, velocity shear is the contributing factor for negative diffusivity, whilst in our case, it is collisionless effects (the Hall term) that play an analogous role. 

\subsection{Where are the Hall effects important?} \label{SSecHImp}
To gain an estimate of the new terms arising from the Hall term, let us focus on the $\alpha$ effect alone. Of the three terms in (\ref{alphaNkin}), only the second one is identical to the MHD case. Carrying out a naive dimensional analysis, we find that 
\begin{equation}
\frac{|{\bf v}_E \cdot \left(\nabla \times {\bf v}_E\right)|}{|{\bf b} \cdot \left(\nabla \times {\bf b}\right)|} \sim \left(\calm_A \pm \epsilon_H\right)^2,
\end{equation}
\begin{equation}
\frac{|{\bf b} \cdot \left(\nabla \times \nabla \times {\bf v}_E\right)|}{|{\bf b} \cdot \left(\nabla \times {\bf b}\right)|} \sim \epsilon_H \left(\calm_A \pm \epsilon_H\right),
\end{equation}
where $\calm_A = v/v_A$ is the turbulent Alfv\'en Mach number and $\epsilon_H = d_i/L$ is the Hall parameter, with $L$ denoting the scale length. The `$\pm$' has been inserted since the velocity and magnetic fluctuations may be parallel or anti-parallel. Allowing for $\epsilon_H \rightarrow 0$ leads us to the equivalent MHD scalings. Thus, the Hall effects become dominant when the Hall parameter is significant, which is entirely along expected lines. Some examples of systems wherein this parameter is large have been presented in Sec. \ref{SecIntro}.

\section{Discussion and Conclusion} \label{SecConc}
In this paper, we have investigated the role of the Hall effect on conventional large scale dynamo theory. For the purposes of physical simplicity, we have used a very simple ansatz, along the lines of \citet{GD94,GD95}, as the basis of our investigations. As discussed in Sec. \ref{SecIntro}, the calculations presented here are only the first step in a more complete treatment of the Hall dynamo, which will be extended in future work(s) to include several effects that have been omitted for simplicity. Foremost among them is the effect of large-scale velocity shear, which has led to the prediction of the novel ``magnetic shear current effect'' \citep{SB16}. We have chosen to work with such a model as it helps clarify the effects engendered by the Hall term. In `real' systems, however, one would expect it to act in concert with all of the other phenomena alluded to in Sec. \ref{SecIntro}.

Our final results are encapsulated by (\ref{betakin}), (\ref{alphaNkin}) and (\ref{alphaquench}). Thus, we find that the $\alpha$ and $\beta$ effects are modified when compared to their MHD counterparts, and that $\alpha$-quenching operates analogous to GD94 and GD95 as well as \citet{VC92,CH96}. Each of these results has important physical interpretations and consequences. We have also carried out a simple dimensional analysis in Sec. \ref{SSecHImp} and concluded that the dynamo coefficients are significantly altered when the Hall parameter is sufficiently large, and the latter does occur in several space, astrophysical and laboratory plasmas, as pointed out in Sec. \ref{SecIntro}.

We observe that the null value of $\alpha$ is achieved via the process of `Beltramization' in Hall MHD, i.e. when the system attains a particular double Beltrami state, as opposed to `Alfv\'enization' \citep{DMV80,GFPL82} in MHD. We find that (i) wave turbulence (mediated via nonlinear Alfv\'en waves), (ii) relaxed states (the double Beltrami states can be thus interpreted), and (iii) large scale dynamo theory are intimately connected in Hall MHD. We believe that the connections between these three (highly important) areas of plasma physics certainly merit further investigation - a similar line of thought has also been advanced in a recent review by \citet{Moff16}, albeit for MHD.

Moreover, there has been a great deal of interest in the diffusivity ($\beta$) becoming negative, and driving the large scale dynamo. Unlike many standard treatments that rely on large scale velocity shear or stochasticity \citep{VB97,RK03,SB16}, we have shown that the isotropic diffusion tensor can become negative under the influence of the Hall term, in concordance with earlier results \citep{MAP07}. We reiterate that the non-positive definiteness of $\beta$ induced by the Hall term is likely to complement (and supplement) the shear-current effect, and not necessarily supplant it.

At this juncture, we proffer a potential explanation as to why $\beta < 0$ is possible. The backreactions of the Lorentz force $\left({\bf J} \times {\bf B}\right)$ are responsible for introducing the current helicity contribution to the $\alpha$ effect that is \emph{opposite} in sign to the kinematic fluid helicity \citep{PFL76,BS05}. However, the same term is also present in the Ohm's law of Hall MHD, and is responsible for new contributions to the dynamo coefficients - this is seen by inspecting the last two terms on the RHS of (\ref{EMFExp}). It is precisely these new contributions that destroy the positive definite nature of $\beta$. Hence, in a manner of speaking, the ${\bf J} \times {\bf B}$ term in the Ohm's law is solely responsible for generating new contributions that `oppose' the tendency for $\beta > 0$. Thus, it is quite plausible that the ${\bf J} \times {\bf B}$ term modifies \emph{both} the $\alpha$ and $\beta$ coefficients and enforces this `opposite' behaviour.

As the Hall term is important in laboratory and astrophysical environments, gives rise to important and subtle physical effects in large scale dynamos, and connects the latter field with other fundamental areas of plasma physics, we suggest that further studies of the kind carried out in the present paper are timely.

\acknowledgments
ML and AB were supported by the NSF Grant No. AGS-1338944 and the DOE Grant No. DE-AC02-09CH-11466. ML is grateful to Pallavi Bhat, Luca Comisso, Fatima Ebrahimi and Nishant K. Singh for their insightful comments and discussions.

\bibliographystyle{aasjournal}
\bibliography{MF-HMHD}

\begin{thebibliography}{}
\expandafter\ifx\csname natexlab\endcsname\relax\def\natexlab#1{#1}\fi

\bibitem[{{Bhattacharjee}(2004)}]{Bhat04}
{Bhattacharjee}, A. 2004, ARA\&A, 42, 365

\bibitem[{{Bhattacharjee} {et~al.}(2001){Bhattacharjee}, {Ma}, \&
  {Wang}}]{BMW01}
{Bhattacharjee}, A., {Ma}, Z.~W., \& {Wang}, X. 2001, Phys. Plasmas, 8, 1829

\bibitem[{{Bhattacharjee} \& {Yuan}(1995)}]{BY95}
{Bhattacharjee}, A., \& {Yuan}, Y. 1995, ApJ, 449, 739

\bibitem[{{Biskamp}(2003)}]{Bisk03}
{Biskamp}, D. 2003, {Magnetohydrodynamic Turbulence} (Cambridge Univ. Press)

\bibitem[{{Blackman}(2003)}]{Black03}
{Blackman}, E.~G. 2003, in Lecture Notes in Physics, Berlin Springer Verlag,
  Vol. 614, Turbulence and Magnetic Fields in Astrophysics, ed. E.~{Falgarone}
  \& T.~{Passot}, 432--463

\bibitem[{{Blackman}(2015)}]{Bla15}
{Blackman}, E.~G. 2015, Space Sci. Rev., 188, 59

\bibitem[{{Blackman} \& {Field}(1999)}]{BF99}
{Blackman}, E.~G., \& {Field}, G.~B. 1999, ApJ, 521, 597

\bibitem[{{Blackman} \& {Field}(2002)}]{BF02}
---. 2002, Phys. Rev. Lett., 89, 265007

\bibitem[{{Boldyrev}(2006)}]{Bol06}
{Boldyrev}, S. 2006, Phys. Rev. Lett., 96, 115002

\bibitem[{{Brandenburg} {et~al.}(2008){Brandenburg}, {R{\"a}dler},
  {Rheinhardt}, \& {K{\"a}pyl{\"a}}}]{Betal08}
{Brandenburg}, A., {R{\"a}dler}, K.-H., {Rheinhardt}, M., \& {K{\"a}pyl{\"a}},
  P.~J. 2008, ApJ, 676, 740

\bibitem[{{Brandenburg} {et~al.}(2012){Brandenburg}, {Sokoloff}, \&
  {Subramanian}}]{BSS12}
{Brandenburg}, A., {Sokoloff}, D., \& {Subramanian}, K. 2012, Space Sci. Rev.,
  169, 123

\bibitem[{{Brandenburg} \& {Subramanian}(2005)}]{BS05}
{Brandenburg}, A., \& {Subramanian}, K. 2005, Phys. Rep, 417, 1

\bibitem[{{Bushby} \& {Tobias}(2007)}]{BT07}
{Bushby}, P.~J., \& {Tobias}, S.~M. 2007, ApJ, 661, 1289

\bibitem[{{Cattaneo} \& {Hughes}(1996)}]{CH96}
{Cattaneo}, F., \& {Hughes}, D.~W. 1996, Phys. Rev. E, 54, R4532

\bibitem[{{Ding} {et~al.}(2004){Ding}, {Brower}, {Craig}, {Deng}, {Fiksel},
  {Mirnov}, {Prager}, {Sarff}, \& {Svidzinski}}]{Det04}
{Ding}, W.~X., {Brower}, D.~L., {Craig}, D., {et~al.} 2004, Phys. Rev. Lett.,
  93, 045002

\bibitem[{{Dobrowolny} {et~al.}(1980){Dobrowolny}, {Mangeney}, \&
  {Veltri}}]{DMV80}
{Dobrowolny}, M., {Mangeney}, A., \& {Veltri}, P. 1980, Phys. Rev. Lett., 45,
  144

\bibitem[{{Ebrahimi} \& {Bhattacharjee}(2014)}]{EB14}
{Ebrahimi}, F., \& {Bhattacharjee}, A. 2014, Phys. Rev. Lett., 112, 125003

\bibitem[{{Galtier} \& {Buchlin}(2007)}]{GB07}
{Galtier}, S., \& {Buchlin}, E. 2007, ApJ, 656, 560

\bibitem[{{G{\'o}mez} {et~al.}(2010){G{\'o}mez}, {Mininni}, \&
  {Dmitruk}}]{GMD10}
{G{\'o}mez}, D.~O., {Mininni}, P.~D., \& {Dmitruk}, P. 2010, Phys. Rev. E, 82,
  036406

\bibitem[{{Grappin} {et~al.}(1982){Grappin}, {Frisch}, {Pouquet}, \&
  {Leorat}}]{GFPL82}
{Grappin}, R., {Frisch}, U., {Pouquet}, A., \& {Leorat}, J. 1982, A\&A, 105, 6

\bibitem[{{Gruzinov} \& {Diamond}(1994)}]{GD94}
{Gruzinov}, A.~V., \& {Diamond}, P.~H. 1994, Phys. Rev. Lett., 72, 1651

\bibitem[{{Gruzinov} \& {Diamond}(1995)}]{GD95}
---. 1995, Phys. Plasmas, 2, 1941

\bibitem[{{Howes} \& {Nielson}(2013)}]{HN13}
{Howes}, G.~G., \& {Nielson}, K.~D. 2013, Phys. Plasmas, 20, 072302

\bibitem[{{Huba}(1995)}]{Huba95}
{Huba}, J.~D. 1995, Phys. Plasmas, 2, 2504

\bibitem[{{Hughes} \& {Proctor}(2009)}]{HP09}
{Hughes}, D.~W., \& {Proctor}, M.~R.~E. 2009, Phys. Rev. Lett., 102, 044501

\bibitem[{{Jackson}(1999)}]{JDJ99}
{Jackson}, J.~D. 1999, {Classical electrodynamics} (Wiley)

\bibitem[{{Ji}(1999)}]{Ji99}
{Ji}, H. 1999, Phys. Rev. Lett., 83, 3198

\bibitem[{{Ji} {et~al.}(1994){Ji}, {Almagri}, {Prager}, \& {Sarff}}]{Ji94}
{Ji}, H., {Almagri}, A.~F., {Prager}, S.~C., \& {Sarff}, J.~S. 1994, Phys. Rev.
  Lett., 73, 668

\bibitem[{{Kleeorin} \& {Rogachevskii}(1994)}]{KR94}
{Kleeorin}, N., \& {Rogachevskii}, I. 1994, Phys. Rev. E, 50, 493

\bibitem[{{Krause} \& {R{\"a}dler}(1980)}]{KR80}
{Krause}, F., \& {R{\"a}dler}, K.-H. 1980, {Mean-field magnetohydrodynamics and
  dynamo theory} (Oxford: Pergamon Press)

\bibitem[{{Kulsrud} \& {Zweibel}(2008)}]{KZ08}
{Kulsrud}, R.~M., \& {Zweibel}, E.~G. 2008, Rep. Prog. Phys., 71, 046901

\bibitem[{{Kunz} \& {Lesur}(2013)}]{KL13}
{Kunz}, M.~W., \& {Lesur}, G. 2013, MNRAS, 434, 2295

\bibitem[{{Lingam} \& {Bhattacharjee}(2016)}]{LB16}
{Lingam}, M., \& {Bhattacharjee}, A. 2016, MNRAS, 460, 478

\bibitem[{{Lingam} \& {Mahajan}(2015)}]{LM15}
{Lingam}, M., \& {Mahajan}, S.~M. 2015, MNRAS, 449, L36

\bibitem[{{Lingam} {et~al.}(2016){Lingam}, {Miloshevich}, \&
  {Morrison}}]{LMM16}
{Lingam}, M., {Miloshevich}, G., \& {Morrison}, P.~J. 2016, Phys. Lett. A, 380,
  2400

\bibitem[{{Lingam} {et~al.}(2015){Lingam}, {Morrison}, \&
  {Miloshevich}}]{LMM15}
{Lingam}, M., {Morrison}, P.~J., \& {Miloshevich}, G. 2015, Phys. Plasmas, 22,
  072111

\bibitem[{{Mahajan} \& {Krishan}(2005)}]{MK05}
{Mahajan}, S.~M., \& {Krishan}, V. 2005, MNRAS, 359, L27

\bibitem[{{Mahajan} \& {Lingam}(2015)}]{ML15}
{Mahajan}, S.~M., \& {Lingam}, M. 2015, Phys. Plasmas, 22, 092123

\bibitem[{{Mahajan} \& {Yoshida}(1998)}]{MY98}
{Mahajan}, S.~M., \& {Yoshida}, Z. 1998, Phys. Rev. Lett., 81, 4863

\bibitem[{{Mininni} {et~al.}(2007){Mininni}, {Alexakis}, \& {Pouquet}}]{MAP07}
{Mininni}, P.~D., {Alexakis}, A., \& {Pouquet}, A. 2007, J. Plasma Phys., 73,
  377

\bibitem[{{Mininni} {et~al.}(2002){Mininni}, {G{\'o}mez}, \& {Mahajan}}]{MGM02}
{Mininni}, P.~D., {G{\'o}mez}, D.~O., \& {Mahajan}, S.~M. 2002, ApJ, 567, L81

\bibitem[{{Mininni} {et~al.}(2003){Mininni}, {G{\'o}mez}, \& {Mahajan}}]{MGM03}
---. 2003, ApJ, 584, 1120

\bibitem[{{Mininni} {et~al.}(2005){Mininni}, {G{\'o}mez}, \& {Mahajan}}]{MGM05}
---. 2005, ApJ, 619, 1019

\bibitem[{{Mirnov} {et~al.}(2003){Mirnov}, {Hegna}, \& {Prager}}]{MHP03}
{Mirnov}, V.~V., {Hegna}, C.~C., \& {Prager}, S.~C. 2003, Plasma Phys. Rep.,
  29, 566

\bibitem[{{Moffatt}(2016)}]{Moff16}
{Moffatt}, H.~K. 2016, Proc. R. Soc. A, 472, 20160183

\bibitem[{{Pouquet} {et~al.}(1976){Pouquet}, {Frisch}, \& {Leorat}}]{PFL76}
{Pouquet}, A., {Frisch}, U., \& {Leorat}, J. 1976, J. Fluid Mech., 77, 321

\bibitem[{{R{\"a}dler} \& {Rheinhardt}(2007)}]{RR07}
{R{\"a}dler}, K.-H., \& {Rheinhardt}, M. 2007, Geophys. Astrophys. Fluid Dyn.,
  101, 117

\bibitem[{{Rogachevskii} \& {Kleeorin}(2003)}]{RK03}
{Rogachevskii}, I., \& {Kleeorin}, N. 2003, Phys. Rev. E, 68, 036301

\bibitem[{{Rogachevskii} \& {Kleeorin}(2004)}]{RK04}
---. 2004, Phys. Rev. E, 70, 046310

\bibitem[{{R{\"u}diger} \& {Kitchatinov}(2006)}]{RK06}
{R{\"u}diger}, G., \& {Kitchatinov}, L.~L. 2006, Astron. Nachr., 327, 298

\bibitem[{{Sano} \& {Stone}(2002)}]{SS02}
{Sano}, T., \& {Stone}, J.~M. 2002, ApJ, 570, 314

\bibitem[{{Squire} \& {Bhattacharjee}(2015)}]{SB15}
{Squire}, J., \& {Bhattacharjee}, A. 2015, Phys. Rev. Lett., 115, 175003

\bibitem[{{Squire} \& {Bhattacharjee}(2016)}]{SB16}
---. 2016, J. Plasma. Phys., 82, 535820201

\bibitem[{{Sridhar} \& {Singh}(2010)}]{SS10}
{Sridhar}, S., \& {Singh}, N.~K. 2010, J. Fluid Mech., 664, 265

\bibitem[{{Subramanian}(2016)}]{Sub16}
{Subramanian}, K. 2016, Rep. Prog. Phys., 79, 076901

\bibitem[{{Subramanian} \& {Brandenburg}(2004)}]{SB04}
{Subramanian}, K., \& {Brandenburg}, A. 2004, Phys. Rev. Lett., 93, 205001

\bibitem[{{Sur} {et~al.}(2007){Sur}, {Shukurov}, \& {Subramanian}}]{SSS07}
{Sur}, S., {Shukurov}, A., \& {Subramanian}, K. 2007, MNRAS, 377, 874

\bibitem[{{Turner}(1986)}]{Turn86}
{Turner}, L. 1986, IEEE Trans. Plasma Sci., 14, 849

\bibitem[{{Vainshtein} \& {Cattaneo}(1992)}]{VC92}
{Vainshtein}, S.~I., \& {Cattaneo}, F. 1992, ApJ, 393, 165

\bibitem[{{Vishniac} \& {Brandenburg}(1997)}]{VB97}
{Vishniac}, E.~T., \& {Brandenburg}, A. 1997, ApJ, 475, 263

\bibitem[{{Vishniac} \& {Cho}(2001)}]{VC01}
{Vishniac}, E.~T., \& {Cho}, J. 2001, ApJ, 550, 752

\bibitem[{{Wardle}(2007)}]{Ward07}
{Wardle}, M. 2007, Ap\&SS, 311, 35

\bibitem[{{Weinberg}(1976)}]{Wein76}
{Weinberg}, S. 1976, Phys. Rev. D, 13, 974

\bibitem[{{Yoshizawa} {et~al.}(2004){Yoshizawa}, {Itoh}, {Itoh}, \&
  {Yokoi}}]{YIIY04}
{Yoshizawa}, A., {Itoh}, S.-I., {Itoh}, K., \& {Yokoi}, N. 2004, Plasma Phys.
  Cont. Fusion, 46, R25

\bibitem[{{Yousef} {et~al.}(2008){Yousef}, {Heinemann}, {Schekochihin},
  {Kleeorin}, {Rogachevskii}, {Iskakov}, {Cowley}, \& {McWilliams}}]{Yet08}
{Yousef}, T.~A., {Heinemann}, T., {Schekochihin}, A.~A., {et~al.} 2008, Phys.
  Rev. Lett., 100, 184501

\end{thebibliography}

\end{document}